\documentclass[times,twocolumn,final]{elsarticle}

\usepackage{framed,multirow}

\usepackage{amssymb}
\usepackage{latexsym}

\usepackage{url}
\usepackage{xcolor}
\definecolor{newcolor}{rgb}{.8,.349,.1}

\usepackage{amsmath,epsfig}
\usepackage{paralist,amssymb, amsfonts}
\usepackage{algorithmic}
\usepackage{algorithm}

\usepackage{tabularx}
\usepackage{setspace}
\usepackage{multirow}

\def\x{{\mathbf x}}

\DeclareMathOperator*{\argmin}{arg\,min}
\DeclareMathOperator*{\argmax}{arg\,max}

\begin{document}

\thispagestyle{empty}

\clearpage
\thispagestyle{empty}
\ifpreprint
  \vspace*{-1pc}
\fi

\clearpage
\thispagestyle{empty}

\ifpreprint
  \vspace*{-1pc}
\else
\fi

\clearpage

\ifpreprint
  \setcounter{page}{1}
\else
  \setcounter{page}{1}
\fi

\begin{frontmatter}

\title{Compressed Sensing ECG using Restricted Boltzmann Machines}

\author[1]{Luisa F. Polan\'{i}a \corref{cor1}}
\cortext[cor1]{Corresponding authors: }
\ead{lpolania@amfam.com}

\author[2]{Rafael I. Plaza \corref{cor1}}
\ead{plaza@udel.edu}

\address[1]{American Family Mutual Insurance Company, Madison, WI 53783, USA}
\address[2]{Scientific Games Corporation, Cedar Falls, IA 50613, USA}


\begin{abstract}
Recently, it has been shown that compressed sensing (CS) has the potential to lower energy consumption in wireless electrocardiogram (ECG) systems. By reducing the number of acquired measurements, the communication burden is decreased and energy is saved. In this paper, we aim at further reducing the number of necessary measurements to achieve faithful reconstruction by exploiting the representational power of restricted Boltzmann machines (RBMs) to model the probability distribution of the sparsity pattern of ECG signals. The motivation for using this approach is to capture the higher-order statistical dependencies between the coefficients of the ECG sparse representation, which in turn, leads to superior reconstruction accuracy and reduction in the number of measurements, as it is shown via experiments.
\end{abstract}


\end{frontmatter}

\section{Introduction}
Even though wireless body area networks (WBANs) have the potential to revolutionize health monitoring by allowing the transition from centralized health care services to ubiquitous and ambulatory health monitoring, they still need to face some challenges, such as reducing energy consumption~\cite{Huas09}. Current Zigbee and ANT-based technologies claim that they have battery life of three years, but this is the case for low operating data rates. For continuous operation at 250 kb/s, a regular Lithium ion battery is consumed in a couple of hours~\cite{Seye13}. Therefore, at the heart of making WBAN ready for deployment is the need for data reduction at the sensor nodes. It has been shown that the application of CS to WBAN-enabled ECG monitors results in power-efficient sensor nodes that attain high compression rates at a low computational cost when using a sparse binary sensing matrix~\cite{Mama11}. Compressed sensing is a framework that exploits the structure of signals to acquire data at a rate proportional to the information content rather than the frequency content, therefore allowing perfect reconstruction from sub-Nyquist random linear measurements~\cite{Dono06,Cand08}.

The application of CS to wireless ECG systems has attracted attention in recent years. Works in this area include studies of practical design considerations for CS encoding and decoding~\cite{Mama11, Dixo12}, as well as the development of CS reconstruction algorithms that exploit structural ECG information~\cite{Zhan13, Mama11a, Pola11, Pant14, Zhan15, Pola15b}. This paper proposes to use our recently developed CS scheme~\cite{Pola17}, henceforth referred to as RBM-CS scheme, for ECG reconstruction. This scheme exploits the representational power of RBMs to model the probability distribution of the sparsity pattern of a signal class, with the ultimate goal of reducing the number of measurements.

In this paper, overcomplete dictionaries and wavelets are employed to sparsely represent ECG signals. Training of the dictionaries also results in a set of sparse codes associated with the training data, which are often discarded after training since the main interest lies in the dictionary. Instead, the RBM-CS scheme utilizes the sparse codes support to train a RBM, which is later used by a CS reconstruction algorithm that fully exploits the model. The reason for incorporating RBMs into the reconstruction is to exploit higher-order dependencies between sparse coefficients. The RBM-CS scheme falls within the structured compressed sensing framework~\cite{Duar11}, which aims at exploiting signal structure by considering more elaborate priors that go beyond the simplistic sparsity prior. Even though ECG signals have a rich structure, most of the previous CS ECG works only exploit signal sparsity. One exception is our recent work that incorporates prior information about the dependencies of the ECG wavelet coefficients across scales into the CS reconstruction algorithm~\cite{Pola15b}. Another work that exploits ECG structural information but using a statistical model was proposed by Zhang \textit{et al}.~\cite{Zhan13}. Their work proposes to use the block sparse Bayesian learning framework to reconstruct fetal ECG recordings.

The performance of the RBM-CS scheme is evaluated on the MIT-BIH Arrhythmia Database~\cite{Gold00} and the European ST-T Database~\cite{Tadd92}. Experimental results include comparisons with the Basis Pursuit denoising algorithm (BPDN),  the most widely used reconstruction algorithm in the CS ECG literature~\cite{Mama11,Dixo12}. Experiments are also performed with the bound-optimization-based BSBL algorithm~\cite{Zhan13}, referred hereafter as BO-BSBL, and with the model-based CoSaMP and model-based Iterative Hard Thresholding algorithms~\cite{Pola15b} in order to compare with other CS reconstruction algorithms that also exploit signal structure. Simulation results indicate that the RBM-CS scheme offers superior reconstruction accuracy for the low-measurement regime.

The structure of the paper is as follows. Section II presents a brief review of RBMs and the RBM-CS scheme. Section III validates the potential of the RBM-CS scheme to improve reconstruction quality and to reduce the number of measurements via experiments. The paper concludes in Section IV with final remarks.

\section{Background}
\subsection{Restricted Boltzmann Machines}
Restricted Boltzmann machines are a type of undirected graphical models formed by a layer of binary stochastic hidden units and a layer of stochastic visible units that, for the purpose of this work, will be Bernoulli distributed conditional on the hidden units. The visible units $v=[v_1 v_2 \ldots v_J]^T$ represent the input variables of the data that needs to be modeled. The hidden units $h=[h_1 h_2 \ldots h_P]^T$ are trained to capture the higher-order correlations that are observed at the visible units. The layers are connected via a weight matrix $W$. The structure of a RBM forms a bipartite graph with no visible-visible or hidden-hidden connections. In this work, RBMs are used to learn a joint probability distribution of training data. In a RBM, the probability distribution over visible units is defined as

\begin{equation}\label{eq6.12}
p({v})=\sum_{{h}} p({v},{h})=-\frac{1}{Z}\text{exp}\left(-\text{E}({v})\right),
\end{equation}

where
\begin{equation}\label{eq6.13}
\text{E}(v)=-\sum_j\text{log}\left(1+e^{W_{\cdot j}^Tv+{b_{h}}_j}\right)-b_v^Tv,
\end{equation}

and $Z=\sum_\mathbf{v} \text{exp}(-\text{E}(\mathbf{v}))$ is the normalization term.

\subsection{Compressed Sensing Scheme using restricted Boltzmann machines}
\begin{figure}[t]
\vspace{-.3cm}
\centering{ 
\includegraphics[width = \columnwidth]{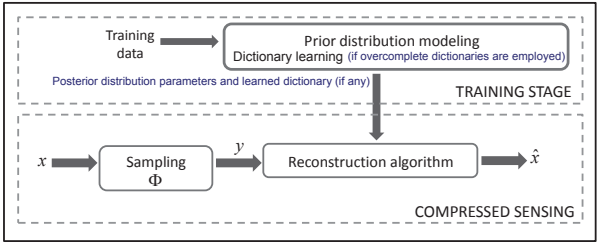}}
\caption{Block diagram of the RBM-CS scheme} \label{block2}
\end{figure}

We recently proposed a CS scheme which employs a RBM to model the probability distribution of the sparsity pattern of a signal class, the RBM-CS scheme~\cite{Pola17}. The main advantage of this scheme is that it captures higher-order dependencies between sparse coefficients, which ultimately translates into a reduction in the number of necessary measurements to attain accurate reconstruction. Fig. \ref{block2} illustrates the block diagram of the RBM-CS scheme, which consists of two stages, namely, the training and compressed sensing stages. The training stage varies depending on the employed sparsifying transform. Details of each stage are provided in this section.

\subsubsection{Compressed Sensing stage}
Let $x\in\mathbb{R}^N$ denote the signal to be recovered and $D\in\mathbb{R}^{N\times Q}$ the dictionary employed to represent $x$, \textit{i.e.} $x=Ds+r$, where $s$ and $r$ are the sparse representation and the representation error. A Gaussian distribution with zero mean and covariance $\Sigma_r$ is assumed for $r$. The support of $s$, of cardinality $K$, is denoted as $\theta$. The nonzero coefficients of $s$ are denoted as $s_\theta$. A Gaussian distribution with zero mean and variance $\sigma_{s_i}^2$ is assumed for each nonzero coefficient $s_i$, $i\in \theta$. Therefore, the conditional distribution of $s_\theta$ given $\theta$ takes the form $s_\theta|\theta \sim \mathcal{N}(\mathbf{0}, \Sigma_\theta)$, where $\Sigma_\theta\in\mathbb{R}^{K\times K}$ is a diagonal matrix, whose diagonal is formed by the variances of the nonzero coefficients $\sigma_{s_i}^2$, $i\in \theta$. The sparsity pattern $S^{\{\theta\}}$ associated with the support $\theta$ is defined as $S_i^{\{\theta\}}=\textbf{1}[i\in \theta]$ for $i=1, \ldots, N$, where $\textbf{1}[\cdot]$ denotes the indicator function.

Compressed sensing addresses the recovery of signal $x$ from undersampled and noisy measurements of the form $y=\Phi x+n$, where $\Phi \in\mathbb{R}^{M\times N}$ is the sampling matrix and $n$ accounts for the additive Gaussian sampling noise of zero mean and variance $\sigma_n^2$. Defining $\Xi=\Phi D$ and $\eta=\Phi r+n$, vector $y$ takes the form $y=\Xi s+\eta$. The conditional distribution of $y$ given $\theta$ is given by

\begin{equation}\label{eq6.10}
\begin{split}
\text{p}(y|\theta)=&C\times{\text{det}\left(\Xi_\theta^T\Sigma_\eta^{-1}\Xi_\theta\Sigma_\theta+I\right)^{-1/2}}\\
   &\times\text{exp}\left\{\frac{1}{2}y^T\Sigma_\eta^{-1}\Xi_\theta P^{-1}\Xi_\theta^T\Sigma_\eta^{-1}y\right\},
\end{split}
\end{equation}
where $C={{\text{det}(2\pi\Sigma_\eta)}^{-1/2}}\text{exp}\left\{-\frac{1}{2}y^T\Sigma_\eta^{-1}y\right\}$ and $P=\Xi_\theta^T\Sigma_\eta^{-1}\Xi_\theta+\Sigma_\theta^{-1}$.

In the RBM-CS scheme, the MAP estimator is employed to recover $s$. The MAP estimate of $s$ requires knowledge of the support, which is calculated as
\begin{eqnarray}\label{eq6.111}
\hat{\theta}&=&\argmax_{\theta} p(\theta|y)\\
&=&\argmax_{\theta} p(y|\theta)p(\theta)\\
&=&\argmax_{\theta} \left(\frac{1}{2}y^T\Sigma_\eta^{-1}\Xi_\theta P^{-1}\Xi_\theta^T\Sigma_\eta^{-1}y    \right.\\
&&  -\frac{1}{2}\text{log}\left(\text{det}(P\Sigma_\theta)\right) +\sum_j\text{log}\left(1+e^{W_{\cdot j}^TS^{\{\theta\}}+{b_{h}}_j}\right) \nonumber\\ 
&& \left.+b_v^TS^{\{\theta\}}\right). \nonumber
\end{eqnarray}
where $p(\theta)$ is calculated by using \eqref{eq6.12}, the probability distribution over the visible units of the RBM. 

The posterior distribution $p(s_{\hat{\theta}}|y, \hat{\theta})$ is Gaussian distributed, and therefore, the MAP estimate of $s$ is directly obtained from the mean of the posterior, \textit{i.e.},
\begin{eqnarray}\label{eq6.18}
\hat{s}_{\hat{\theta}}&=&\argmax_{s_{\hat{\theta}}} p(s_{\hat{\theta}}|y, \hat{\theta}),\\
&=&\Sigma_{\hat{\theta}} \Xi_{\hat{\theta}}^T(\Xi_{\hat{\theta}}\Sigma_{\hat{\theta}} \Xi_{\hat{\theta}}^T+\Sigma_\eta)^{-1}y. \nonumber
\end{eqnarray}

To solve for \eqref{eq6.111}, the RBM-CS scheme uses a greedy pursuit algorithm based on the Orthogonal Matching Pursuit (OMP) algorithm, referred to as RBM-OMP-like algorithm~\cite{Pola17}. The algorithm starts by initializing the support to the empty set. Then, it searches for the element ${i}$ that can be added to the support in order to maximize $p(\theta|y)$ at each iteration. The algorithm stops when the number of iterations exceeds the pre-defined sparsity threshold. Once the signal support is calculated, the sparse representation $s$ is estimated via the MAP estimator \eqref{eq6.18}.

\subsubsection{Training stage}
\label{overcomplete}
 Let us first consider the case when overcomplete dictionaries are used as the sparsifying transform. In this case, the training stage is employed with dual purpose, it learns both the the dictionary and the parameters of the posterior distribution $p(\theta|y)$.

Let $G=[g^{1} \ldots g^{B}] \in \mathbb{R}^{N\times B}$ denote the set of $N$-dimensional training samples, which is referred to as training data set. The overcomplete dictionary $D=[d^{1} \ldots d^{J}] \in \mathbb{R}^{N\times J}$ $\left(J>N\right)$ is learned by solving the following optimization problem
\begin{equation}\label{eq6.6}
\{\hat{D}, \hat{A}\}=\argmin_{D, A} \|G-DA\|_{F}^2~~\textrm{s.~t.}~~\|a^{i}\|_{0}<K,~~\forall i,
\end{equation}
where $A=[a^{1} \ldots a^{B}] \in \mathbb{R}^{J\times B}$ and $K$ denote the sparse codes of $G$ and the pre-specified sparsity threshold, respectively. The representation error is defined as $E=G-\hat{D}\hat{A}$. The RBM-CS scheme uses the K-SVD algorithm proposed by Aharon \textit{et al.}~\cite{Ahar06} to solve for~\eqref{eq6.6}.

Let $u^{i}$ denote the sparsity pattern of the sparse code $a^{i}$, $i=1, \ldots, B$. The $j$th element of $u^{i}$ is defined as $u^{i}_j=\textbf{1}[j \in \text{supp}(a^{i})]$. The set of vectors  $U=[u^{1} \ldots u^{B}]$ are employed to train the RBM model. In the RBM-CS scheme, contrastive divergence~\cite{Hint02} is used for learning the parameters of the RBM model. Furthermore, the set of  sparse codes $A$ are employed to learn the variances of the sparse coefficients:
\begin{equation}\label{eq6.7}
\hat{\sigma}_{s_i}^2=\frac{\sum_{j=1}^B {\left(a^{j}_i\right)}^2}{\sum_{j=1}^B \textbf{1}[i\in \text{supp}\left(a^{j}\right)]}, i=1\ldots N
\end{equation}

The RBM-CS scheme assumes that the sampling noise variance is known \textit{a priori}. It also assumes independence between the representation error coefficients $r_i$ and $r_j$ for $i\neq j$, which implies that the the covariance matrix $\Sigma_r$ is a diagonal matrix, whose diagonal is formed by the variances of the representation error coefficients $\sigma_{r_i}^2$, $\forall i$. The representation error of the learned dictionary $E=[e^{1} \ldots e^{B}]$ is employed to estimate each diagonal element of $\Sigma_r$. That is,
\begin{equation}\label{luisilla}
\hat{\sigma}_{r_i}^2=\frac{1}{B} \sum_{j=1}^B \left(e_{i}^{j}\right)^2.
\end{equation}
The estimate of $\Sigma_\eta$ is directly calculated as $\hat{\Sigma}_\eta=\Phi\hat{\Sigma}_r\Phi^T+\sigma_n^2I$.
In the case when the sparsifying transform is an orthonormal basis, the training stage is only used to learn the parameters of the posterior distribution $p(\theta|y)$, following the same procedure as in the case of overcomplete dictionaries.

\section{Experimental Results}
To validate the potential of the RBM-CS scheme to improve the performance of CS ECG systems, experiments are performed on records from the MIT-BIH Arrhythmia Database~\cite{Gold00} and the European ST-T database~\cite{Tadd92}. 

The entries of the sampling matrix $\Phi$ are independently sampled from a symmetric Bernoulli distribution (P($\Phi_{i,j}=\pm1/\sqrt{M}=1/2$)) in order to build an efficient hardware implementation. The use of Bernoulli matrices, as compared to other sub-Gaussian
matrices, results in simpler circuit complexity, data storage, and computation requirements~\cite{Chen12}. Presented results correspond to averages of 50 repetitions of each experiment, with a different realization of the random measurement matrix at each time. Both the wavelet transform and learned overcomplete dictionaries are employed as sparsifying transforms in the experiments. 

The reconstruction SNR (R-SNR), precision, recall, and the percentage similarity are used as performance measures for our experiments. The R-SNR is defined as
\begin{equation}\label{BPD1807}
\text{R-SNR}=10\text{log}_{10}\frac{\|x\|_2^2}{\|x-\hat{x}\|_2^2},
\end{equation}
where $x$ and $\hat{x}$ denote the $N$-dimensional original and reconstructed signals, respectively. Precision is defined as $Precision=\text{TP}/(\text{TP}+\text{FP})$ and recall is defined as $Recall=\text{TP}/(\text{TP}+\text{FN})$, where TP, FP, and FN denote true positives, false positives and false negatives, respectively. The percentage similarity, also used in~\cite{Crav15b}, is defined as $\text{PSim}=100-\left(\frac{|y-\bar{y}|}{y}\times 100\right)$, where $y$ is the value of a calculated metric using the original signal and  $\bar{y}$ is the value of the same metric on the resulting signal after reconstruction. PSim is used for the precision and recall metrics in this paper. Precision, recall, and the percentage similarity are used for the performance evaluation of QRS detection in Section \ref{sec:QRS}.

\begin{figure}[t]
\centering{ 
\includegraphics[width = \columnwidth]{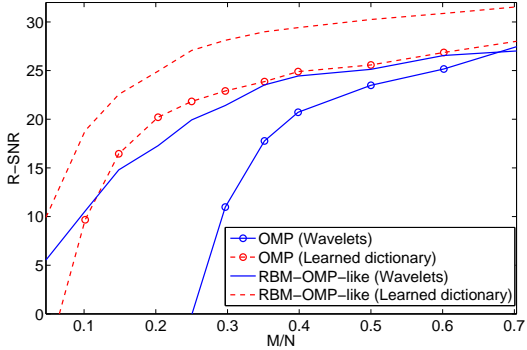}}
\caption{Comparison of the reconstruction of ECG signals using the OMP and the RBM-OMP-like algorithm for both wavelets and learned overcomplete dictionaries.} \label{Fig1}
\end{figure}

\subsection{Experiments on the MIT-BIH Arrhythmia Database}
Every file in the MIT-BIH Arrhythmia Database contains two 30-min long lead recordings sampled at 360 Hz with 11 bits per sample of resolution. Single leads from records 100, 101, 102, 107, 109, 111, 115, 117, 118 and 119 are employed for the experiments in this section. This data set consists of a variety of signals with different morphologies, rhythms and abnormal heartbeats.

\subsubsection{Performance Evaluation}
\label{sec:performance}
The first experiment compares the ECG reconstruction performance of the RBM-OMP-like algorithm with the traditional OMP algorithm. The training and testing data sets for this experiment consist of 92800 and 5000 segments of size $N=128$, extracted from the selected ECG records from the MIT-BIH Arrhythmia Database, respectively. The same number of segments is extracted from each record. A RBM with the same number of hidden units as of visible units is employed to model the probability distribution of the sparsity pattern. Compressed measurements are artificially contaminated with Gaussian noise of variance $\sigma_n^2=0.25$. The number of dictionary atoms is set to $3N$. The Daubechies-4 wavelet transform, using a decomposition level $L=4$, is employed. The sparsity threshold is set to $K=0.08N$ and $K=0.1N$ for overcomplete dictionaries and wavelets, respectively. The results of the comparison in Fig.~\ref{Fig1} indicate that the RBM-OMP-like algorithm has superior reconstruction performance than OMP and requires significantly less number of measurements to achieve accurate reconstruction. The performance gain is larger in the case of overcomplete dictionaries than in the case of wavelets.

\begin{figure}[t]
\centering{ 
\includegraphics[width = \columnwidth]{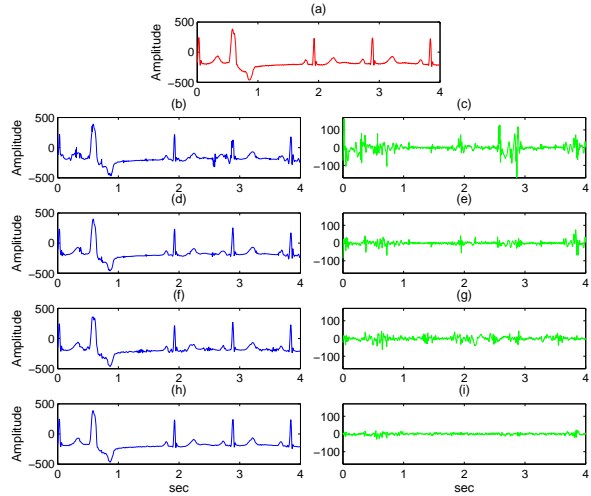}}
\caption{Visual evaluation of the RBM-OMP-like reconstruction algorithm using overcomplete dictionaries and the wavelet basis as sparsifying transforms. Record 119, $M=0.35N$. First row: (a) Original signal. Second and third rows: (b-e) Reconstructed signals using a wavelet basis as the sparsifying transform (b-c) OMP reconstruction and error, R-SNR=18.32, (d-e) RBM-OMP-like reconstruction and error, R-SNR=25. Fourth and fifth rows: (f-i) Reconstructed signals using an overcomplete dictionary as the sparsifying transform (f-g) OMP reconstruction and error, R-SNR=24.59, (h-i) RBM-OMP-like reconstruction and error, R-SNR=31.67.} \label{Fig2}
\end{figure}

Fig.~\ref{Fig2} visually illustrates the reconstruction of an ECG signal using the RBM-OMP-like algorithm. Fig.~\ref{Fig2}(a) corresponds to a 4-second duration segment from record 119, which contains ventricular ectopic beats. The signal is divided into segments of 128 samples. Each segment is sampled and reconstructed separately. Then, the segments are concatenated to reconstruct the original signal. For this experiment, the number of measurements is set to $M = 0.35N$. Figs.~\ref{Fig2}(b-e) and Figs.~\ref{Fig2}(f-i) are obtained when wavelets and overcomplete dictionaries are employed as the sparsifying transform, respectively. Figs.~\ref{Fig2}(b-c) and Figs.~\ref{Fig2}(f-g) show the obtained reconstructed (left) and error (right) signals when using the OMP algorithm. Similarly, Figs.~\ref{Fig2}(d-e) and Figs.~\ref{Fig2}(h-i) show the obtained reconstructed (left) and error (right) signals when using the RBM-OMP-like algorithm.
The recovered signals using the RBM-OMP-like algorithm are better estimates of the original signals than those obtained with the traditional OMP algorithm. It should be noted that using overcomplete dictionaries leads to preservation of detailed information for clinical diagnosis and less number of artifacts in the reconstruction.

\begin{figure}[t]
\centering{ 
\includegraphics[width = \columnwidth]{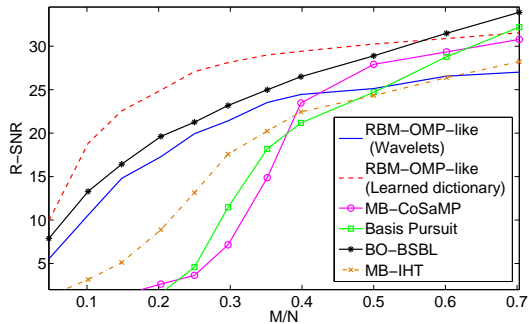}}
\caption{Comparison of the reconstruction of ECG signals using the RBM-OMP-like algorithm, Basis Pursuit Denoising, MB-CoSaMP, MB-IHT, and the BO-BSBL algorithm.} \label{Fig3}
\end{figure}

The second experiment aims at comparing the performance of the RBM-OMP-like algorithm with previously applied CS algorithms to the problem of ECG reconstruction, such as BPDN~\cite{Mama11,Dixo12}, model-based CoSaMP, denoted as MB-CoSaMP, model-based Iterative Hard Thresholding, denoted as MB-IHT~\cite{Pola15b}, and the BO-BSBL algorithm~\cite{Zhan13}. Results are shown in Fig.~\ref{Fig3}. RBM-OMP-like algorithm is the only algorithm that uses both wavelets and overcomplete dictionaries in this experiment. The other algorithms only use wavelets as the sparsifying transform. The second experiment uses the same training and testing datasets of the first experiment.

The most commonly used reconstruction algorithm in the CS ECG literature is BPDN. MB-CoSaMP and MB-IHT exploit the connected subtree structure formed by the largest (in magnitude) wavelet coefficients. The BO-BSBL algorithm was previously employed to reconstruct non-invasive fetal ECG~\cite{Zhan13}, however, it can also be successfully applied to the recovery of adult ECG due to the clustering property of the ECG wavelet coefficients that motivates the use of the block-sparsity model assumed in the BO-BSBL algorithm. For the low-measurement range ($M<0.4N$), the RBM-OMP-like algorithm using wavelets outperforms MB-CoSaMP, MB-IHT, and the BPDN algorithms. The RBM-OMP-like algorithm using an overcomplete dictionary exhibits the best performance for $M<0.55N$. It is only outperformed by the BO-BSBL algorithm when a large number of measurements are available, which is not the case of interest for WBAN-enabled ECG monitors. The IEEE 802.15.4 MAC protocol (Zigbee) is the standard for WBANs and is able to operate at 250 kbps (2.4 GHz), 40 kbps (915 MHz) and 20 kbps (868 MHz)~\cite{Seye13}. The data rate of 20 kbps is of special interest since it leads to low energy consumption and extended battery life. In~\cite{Chav09}, it was reported that WBAN-enabled ECG typically have a raw data rate of 10-100 kbps. To satisfy the upper limit of 100 kbps with Zigbee operating at 20 kbps, $M/N=0.2$ would be needed. 

\subsubsection{Qualitative Observations}
\begin{figure}[t]
\centering{ 
\includegraphics[width =.8\columnwidth]{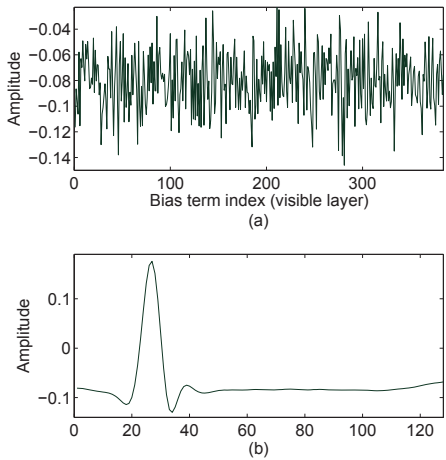}}
\caption{(a) Visible layer bias terms of the RBM model using learned overcomplete dictionaries. (b) Dictionary atom sharing the same index as the largest visible bias term.}\label{Fig9}
\end{figure}

This section presents a qualitative assessment of the trained RBM models. Fig.~\ref{Fig9}(a) illustrates the visible layer bias terms of the RBM using overcomplete dictionaries. Note that the bias terms are negative since the elements of the sparse representation are zero most of the time. Fig.~\ref{Fig9}(b) illustrates the dictionary atom sharing the same index as the largest bias term. Note the resemblance of the dictionary atom with the QRS complex of an ECG cycle. 

\begin{figure}[t]
\centering{ 
\includegraphics[width =.9\columnwidth]{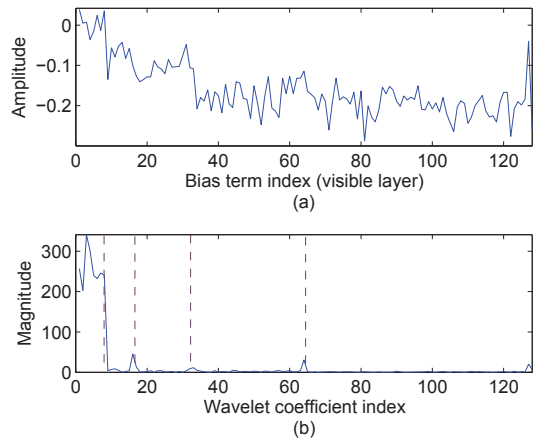}}
\caption{(a) Visible layer bias terms of the RBM model using wavelets. (b) Wavelet representation of an ECG segment of 128 samples.}\label{Fig8}
\end{figure}

Fig.~\ref{Fig8}(a) illustrates the visible layer bias terms of the RBM using wavelets. Most of the bias terms are negative. Fig.~\ref{Fig8}(b) illustrates the magnitude of the wavelet representation of a segment from record 117, where the dashed lines separate each wavelet subband. Sparse representation coefficients sharing the same indexes as the  largest bias terms of the visible layer are likely to belong to the best $K$-term sparse approximation. This is consistent with the result in Fig.~\ref{Fig8}(a) showing that the largest bias term indexes correspond to the indexes of the scaling wavelet coefficients (8 first indexes of the wavelet representation), which accumulate most of the ECG signal energy~\cite{Pola11}, and therefore, are likely to belong to the best $K$-term sparse approximation. Note that the magnitude of the wavelet coefficients tends to decay across scales. A similar behavior is observed for the bias terms of the visible layer (Fig.~\ref{Fig8}(a)). The detail coefficients of the ECG wavelet representation at the highest level are often disregarded as they are expected to have low magnitude~\cite{Shar14, Pola15b}. However, Fig.~\ref{Fig8}(a) shows that the 127th bias term, corresponding (in location) to the highest-level wavelet subband, has a large magnitude, which suggests that the wavelet coefficient with index 127 has a high \textit{a priori} likelihood of belonging to the signal support.

\begin{figure}[t]
\centering{ 
\includegraphics[width =.9\columnwidth]{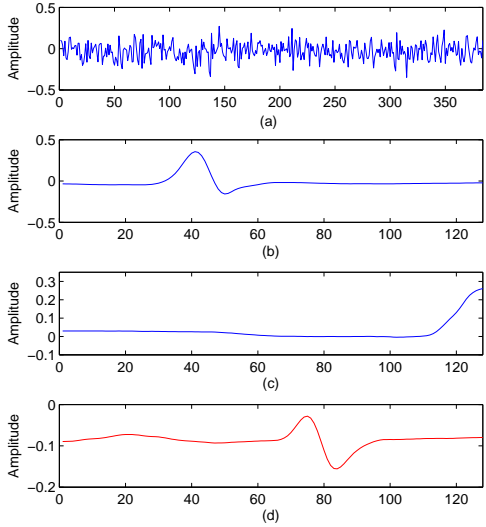}}
\caption{(a) Weights associated with the hidden unit with the largest bias term. (b-c) Dictionary atoms sharing the same indexes as the visible units with the 2 most positive weights. (d) Dictionary atom sharing the same index as the visible unit with the most negative weight.}\label{Fig10}
\end{figure}

The weight matrix of the RBM model reveals coefficient dependencies. Let $h_\star$ denote the hidden unit with the largest bias term of the RBM model using the learned overcomplete dictionary. Fig.~\ref{Fig10}(a) shows the weights associated with $h_\star$. This hidden unit is connected to a group of visible units via large positive weights, which tend to be active simultaneously with $h_\star$, in order to lower the system energy. Figs.~\ref{Fig10}(b) and (c) show the dictionary atoms sharing the same indexes as the 2 visible units with the most positive connections to $h_\star$. Contrarily, Fig.~\ref{Fig10}(d) corresponds to the dictionary atom that shares the same index as the visible unit connected to $h_\star$ via the most negative weight. Note that both the patterns in Figs.~\ref{Fig10}(b) and (d) resemble the QRS complex, and therefore, are unlikely to happen simultaneously in an ECG segment of 128 samples (sampling frequency of 360 Hz).

\begin{figure}[t]
\centering{ 
\includegraphics[width = 1.07\columnwidth]{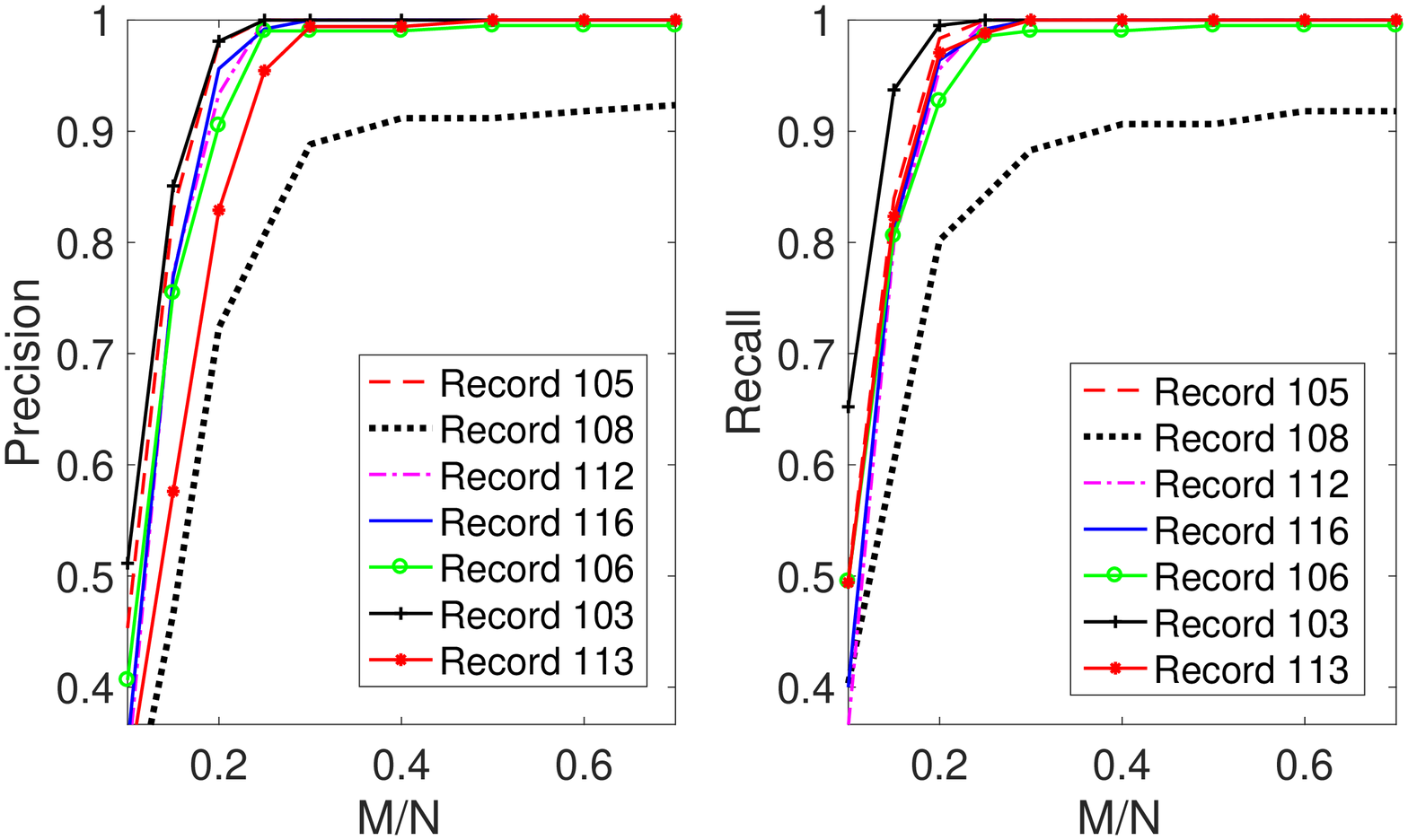}}
\caption{Precision and recall performance for the detection of QRS complexes.} \label{figure_precision}
\end{figure}

\subsubsection{QRS Detection Performance}
\label{sec:QRS}

The transmission and reconstruction of ECG signals is typically followed by other processing tasks, \textit{e.g.} arrhythmia classification, whose first step is usually the detection of the QRS complexes. Therefore, it is worth evaluating how the performance of a QRS detection algorithm varies with the number of measurements. To achieve this goal, the threshold-independent QRS detector proposed in~\cite{Rama14} is used. The ground truth for the R peaks is determined by the QRS detection provided by the algorithm when using the original ECG signal as input. The QRS detection performance of the reconstructed signals is measured using precision and recall. An R-peak detection in the reconstructed signal is classified as true positive when it is within $\pm 4$ samples from an R-peak detected in the original signal. Otherwise, it is classified as false positive. For example, if an R-peak is detected in the original signal at sample 200, a true positive detection in the reconstructed signal will be obtained if an R-peak is detected anywhere in the sample range [196, 204]. If none R-peaks in the reconstructed signal are detected within $\pm 4$ samples from an R-peak detected in the original signal, it will lead to a false negative result. 

The same RBM model trained in Section~\ref{sec:performance} is employed for reconstruction in this experiment. Measurements are artificially contaminated with Gaussian noise of variance $\sigma_n^2=0.1$. The Daubechies-4 wavelet transform, using a decomposition level $L=4$, and a sparsity threshold $K=0.1N$ is employed as well. Records 103, 105, 106, 108, 112, 113 and 116 are selected to evaluate the performance of the QRS detector since they were not employed to train the RBM in Section~\ref{sec:performance}. The selected ECG records are first sampled using a certain number of measurements and a sliding window of size $N=128$, then reconstructed using the RBM-OMP-like algorithm. Finally, the reconstructed segments of length $N=128$ are concatenated to reconstruct the original signal and the R-peaks are detected. Precision and recall results as a function of the number of measurements are presented in Fig. \ref{figure_precision}. High precision values, above 0.8, are obtained even for low number of measurements $M=0.25N$ for all the selected ECG records. Similarly, high recall values, above 0.8, are obtained for $M \geq 0.2N$.

As a baseline for comparison, QRS detection is also performed with signals reconstructed using the BPDN algorithm. The same subset of ECG records, noise variance, RBM model, wavelet transform settings, sliding window size, and sparsity threshold used in the previous experiment are employed for the comparison. Precision and recall values are averaged across the selected ECG records for both the proposed algorithm and the BPDN algorithm. The results, shown in Table \ref{table_example}, indicate that the RBM-OMP-like algorithm significantly outperforms the BPDN algorithm in terms of QRS detection performance.

\begin{table*}[t]
\footnotesize
\renewcommand{\arraystretch}{1.3}
\caption{Comparison of the RBM-OMP-like algorithm with the BPDN algorithm in terms of QRS detection performance.}
\label{table_example}
\centering
\begin{tabular*}{0.85\textwidth}{@{\extracolsep\fill}c|c|cccccccc}
\hline
\multirow{2}{2cm}{\centering Algorithm}&\multirow{2}{2cm}{\centering Metric}&\multicolumn{8}{c}{$M/N$}\\
\cline{3-10}
 &&0.1&0.15&0.2&0.25&0.3& 0.4 & 0.5 & 0.6\\
\hline
\multirow{3}{2.3cm}{\centering RBM-OMP-like}& Precision& 0.382& 0.746& 0.9& 0.976& 0.982& 0.989 & 0.988 & 0.988\\
&Recall& 0.465& 0.833& 0.937& 0.981& 0.983& 0.989& 0.988 & 0.988\\
\hline
\multirow{3}{2.3cm}{\centering BPDN}& Precision & 0.275 & 0.607 & 0.866 & 0.937 & 0.974& 0.984& 0.988 & 0.988\\
&Recall & 0.338 & 0.68 & 0.911 & 0.967  & 0.977 & 0.984  & 0.987 & 0.987 \\
\hline
\end{tabular*}
\end{table*}

To the best of our knowledge, only the works in~\cite{Duar11} and~\cite{Crav15b} address the problem of QRS detection using signals reconstructed with compressed sensing methods. However, the authors in~\cite{Duar11} only reported QRS detection performance as a function of the number of measurements, but did not report the length of the signals. Without that information, it is not possible to estimate the $M/N$ ratio and compare~\cite{Duar11} with our proposed method. The work in~\cite{Crav15b} reported PSim values for precision and recall above $62\%$ for a compression ratio of $N/M=25$, using a sliding window of size 1024 and overcomplete patient-agnostic dictionaries. By using the proposed algorithm with overcomplete dictionaries, the same sliding window as in~\cite{Crav15b}, and a compression ratio of $N/M=30$, PSim values for precision and recall of 53\% and 59\% are obtained. The percentage of similarity metric depends on the employed QRS detection algorithm. A QRS detector different from that used in~\cite{Crav15b} is employed in this paper, which may have contributed to the difference in performance. In addition, the results in~\cite{Crav15b} were obtained with all the records from the MIT-BIH Arrhythmia Database, while our results correspond to the selected subset of ECG records. 

The work in~\cite{Elge17} evaluates QRS detection on reconstructed signals after using compression methods. A comparison with~\cite{Elge17} is not entirely fair since compressed sensing is not a traditional compression method. Compressed sensing needs to use the compressed measurements to estimate not only the magnitude of the most significant coefficients, but also their positions. As a reference, it is reported that the work in~\cite{Elge17} detects the QRS of reconstructed signals, which were previously compressed with a compression ratio of 4.5, with recall and precision values of $99.78\%$ and $99.92\%$, respectively. They used all the 48 records from the MIT-BIH Arrhythmia Database to validate performance. Using our proposed method with a ratio $N/M=4.5$ leads to averaged recall and precision values of $96.34\%$ and $93.92\%$, respectively, across records 103, 105, 106, 108, 112, 113 and 116. The metric values of the proposed method are smaller because, as indicated before, compressed sensing is not a traditional compression method, and it has been shown that it does not compare favorably with compression methods when only compression ratio is considered~\cite{Crav15}. The value of compressed sensing is that it allows sub-Nyquist sampling. In addition, unlike the work in~\cite{Elge17}, the complete set of records from the MIT-BIH Arrhythmia Database was not used for testing in this experiment because some of the records had already been used for training.

\subsubsection{Robustness to Noise}
\begin{figure}[t]
\centering{ 
\includegraphics[width = \columnwidth]{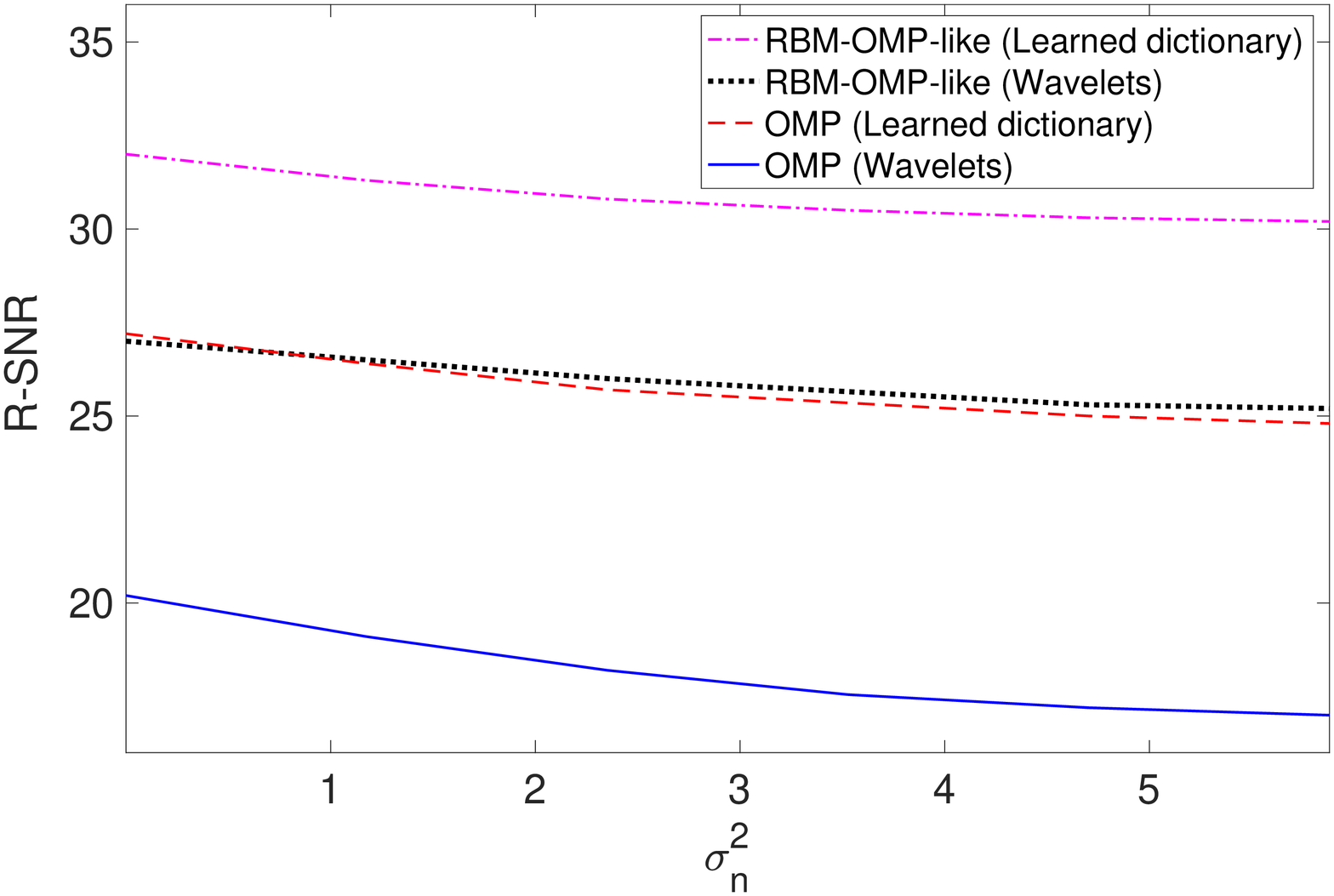}}
\caption{Performance of the RBM-OMP-like and OMP algorithms in the presence of measurement noise.} \label{figure_noise}
\end{figure}

The third experiment evaluates the robustness of the RBM-OMP-like algorithm to measurement noise. An example of measurement noise is quantization noise. Even though it was assumed that the measurement noise was Gaussian in Section~\ref{sec:performance}, it is assumed to be uniformly distributed in this section to simplify the calculations. Using this assumption, the noise variance of an uniform quantizer takes the form $\sigma_n^2={(\Delta f)^2}/{(12\times 2^{2m})}$, where $\Delta f$ is the difference between the maximum and minimum values of the signal to be quantized and $m$ is the number of bits.  Since analog-to-digital converters of 10 bits are often employed in WBANs~\cite{Thot14}, the number of bits $m$ is set to 10. Since $\Delta f$ for the compressed measurements calculated in Section~\ref{sec:performance} ranges from 37.5 to 8600, the noise variance varies approximately from $1\times 10^{-4}$ to 5.88. Therefore, for this experiment, all the parameters are set to the same values as in the first experiment, except the measurement noise variance, which is varied in the range $[1\times 10^{-4}, 5.88]$, while the  Gaussian sensing matrix is kept fixed. Results are shown in Fig.~\ref{figure_noise}. The successful performance of the RBM-OMP-like algorithm in the presence of sampling noise is expected as the noise variance is explicitly taken into account in the algorithm.

\subsubsection{Parameter Evaluation}
\begin{figure}[t]
\centering{ 
\includegraphics[width = \columnwidth]{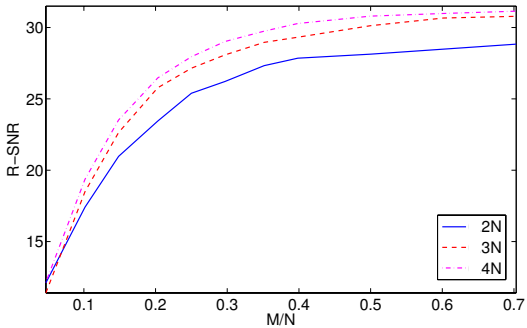}}
\caption{Reconstruction performance of the RBM-OMP-like algorithm using an overcomplete dictionary as a function of the dictionary size.} \label{Fig6}
\end{figure}

We evaluate the sensitivity of the RBM-OMP-like algorithm using overcomplete dictionaries to other parameters, such as the  dictionary size and the sparsity threshold. In each experiment, we only changed the values of one parameter and kept the other parameters fixed. The default values are set the same as in the first experiment. Fig.~\ref{Fig6} illustrates the performance of the RBM-OMP-like algorithm for dictionaries of size 2N, 3N and 4N. The increase in the number of atoms improves the capability of capturing details and thus leads to better representations.

\begin{figure}[t]
\centering{ 
\includegraphics[width = \columnwidth]{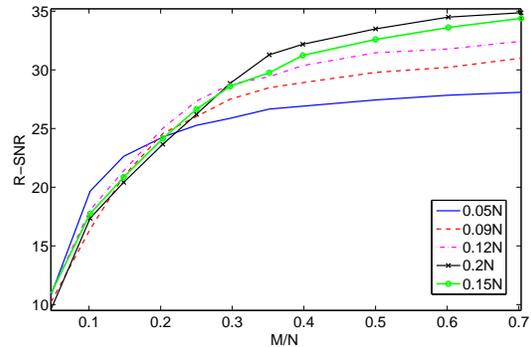}}
\caption{Reconstruction performance of the RBM-OMP-like algorithm using an overcomplete dictionary as a function of the sparsity threshold.} \label{Fig5}
\end{figure}

Fig.~\ref{Fig5} illustrates how the sparsity threshold affects the performance of the RBM-OMP-like algorithm when overcomplete dictionaries are employed as the sparsifying transform. For $M>0.3N$, the R-SNR decreases rapidly as the sparsity threshold decreases from 0.2N to 0.05N. The reason is that useful information is lost when the sparsity threshold decreases. On the other hand, for $M<0.2N$ the R-SNR tends to be slightly higher for smaller sparsity thresholds since sparser signals require fewer measurements to yield stable reconstruction~\cite{Cand08}.

These results demonstrate that the RBM-OMP-like algorithm is not extremely sensitive to variations of the selected parameters. Even though we are aware that optimizing the parameters may lead to reconstruction improvement, how to perform the optimization is still an open problem that is beyond the scope of this paper.

\subsection{Experiments on the European ST-T Database}
The European ST-T Database consists of 90 annotated excerpts of ambulatory ECG recordings from 79 subjects. Each record is two hours long and contains two signals, each sampled at 250 samples per second with 12-bit resolution over a nominal 20 millivolt input range. Single leads from records 103, 104, 106, 107, 108, 110, 112, 118, 123, 129, 136, and 147 are employed for the experiments in this section. This data set consists of a variety of signals with changes in ST segment and T-wave morphology, rhythm, and signal quality.

Experiments on the European ST-T Database evaluate the performance of the RBM-OMP-like algorithm by varying the number of hidden units of the RBM employed to model the probability distribution of the sparsity pattern. The training and testing data sets for this experiment consist of 68000 and 16372 segments of size $N=256$, extracted from the selected ECG records from the European ST-T Database, respectively. The same number of segments is extracted from each record. Compressed measurements are artificially contaminated with Gaussian noise of variance $\sigma_n^2=0.4$. 

\begin{figure}[t]
\centering{ 
\includegraphics[width =0.985 \columnwidth]{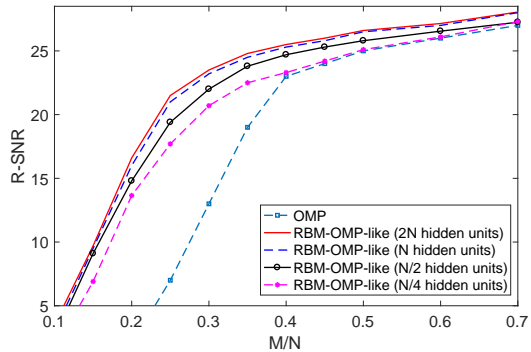}}
\caption{Evaluation of the reconstruction of the selected records from the European ST-T Database using RBM models with different number of hidden units and the wavelet transform as the sparsifying transform.} \label{Fig_8}
\end{figure}

\begin{figure}[t]
\centering{ 
\includegraphics[width =0.97\columnwidth]{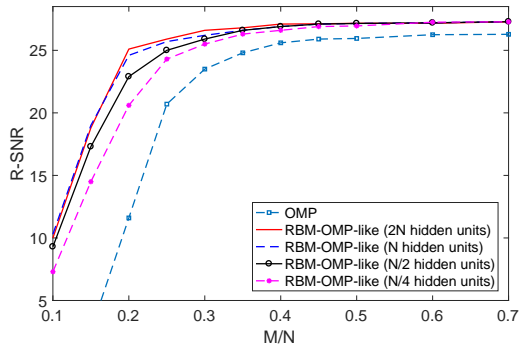}}
\caption{Evaluation of the reconstruction of the selected records from the European ST-T Database using RBM models with different number of hidden units and a learned overcomplete dictionary as the sparsifying transform.} \label{Fig_9}
\end{figure}

Figures~\ref{Fig_8} and \ref{Fig_9} illustrate the performance of the RBM-OMP-like algorithm while varying the number of hidden units of the RBM model and using wavelets and learned overcomplete dictionaries as the sparsifying transform, respectively. The Daubechies-4 wavelet transform, using a decomposition level $L=4$ and a sparsity threshold $K=0.1N$, was employed to generate the results in Fig.~\ref{Fig_8}. Similarly, a learned overcomplete dictionary with $3N$ dictionary atoms and a sparsity level $K=0.12N$ is employed to generate the results in Fig.~\ref{Fig_9}. The mean R-SNR across samples of the testing dataset increases as the number of hidden units changes from $P=N/4$ to $P=N$, respectively. Setting the number of hidden units to $P=2N$ improves the reconstruction performance slightly compared to the case when $P=N$.

\section{Conclusions}

The potential of CS to lower energy consumption in WBAN-enabled ECG monitors lies in its ability to reduce the number of samples at the sensor node. With the goal of further reducing the number of samples, we proposed to use the RBM-CS scheme for ECG reconstruction. This scheme fully exploits signal structure in a statistical fashion, and subsequently, leads to stable reconstruction with low number of measurements. Even though other CS algorithms that exploit signal structure have been applied to ECG reconstruction, they all assume a specific type of structure, \textit{e.g.}, tree-structured sparsity model and block-sparsity model, which is also tied to a specific sparsifying transform. Instead, the RBM-OMP like algorithm uses a general model that accounts for a large set of structures and sparsifying transforms. It was shown that significant performance gains in the low-measurement regime can be obtained by using the RBM-OMP-like algorithm relative to state-of-the-art algorithms for ECG reconstruction. It was also shown that the algorithm is robust to quantization measurement noise and that the detection of QRS complexes on the reconstructed signals remains accurate for $M/N \geq 0.25$. Simulations using real ECG data from the MIT-BIH Arrhythmia database and the the European ST-T database revealed that the performance of the algorithm is better when using learned overcomplete dictionaries than when using wavelets.

\bibliographystyle{elsarticle-num}
\bibliography{refs}

\end{document}